\documentclass{nature_mod}
\usepackage{graphicx}
\usepackage{graphics}
\usepackage{ulem}

\usepackage{amsmath,amssymb}
\newcommand\kms{{\rm\,km\,s^{-1}}}
\newcommand\msun{\rm\,M_\odot}
\newcommand\lsun{\rm\,L_\odot}
\newcommand\rsun{\rm\,R_\odot}
\newcommand\hii{H\,{\sc ii} \,}
\newcommand\myr{\msun \, {\rm yr}^{-1}}

\title{A massive white-dwarf merger product prior to collapse}

\author{Vasilii V.\ Gvaramadze$^{1,2,3}$, G\"otz\ Gr\"afener$^{4}$, Norbert\ Langer$^{4,5}$, Olga V.\ Maryeva$^{6,1}$, \\
Alexei Y.\ Kniazev$^{7,8,1}$, Alexander S.\ Moskvitin$^9$ \& Olga I.\ Spiridonova$^9$}

\begin{document}

\maketitle

{
\begin{affiliations}
\item Sternberg Astronomical Institute, Lomonosov Moscow State University, Universitetskij Pr.~13, Moscow 119992, Russia
\item Space Research Institute, Russian Academy of Sciences, Profsoyuznaya 84/32, 117997 Moscow, Russia
\item Isaac Newton Institute of Chile, Moscow Branch, Universitetskij Pr.\ 13, Moscow 119992, Russia
\item Argelander-Institut f\"ur Astronomie, Universit\"at Bonn, Auf dem H\"{u}gel 71, 53121 Bonn, Germany
\item Max-Planck-Institut f\"ur Radioastronomie, Auf dem H\"{u}gel 69, 53121 Bonn, Germany
\item Astronomick\'y \'ustav, Akademie v\v{e}d \v{C}esk\'e Republiky, Fri\v{c}ova 298, 251 65 Ond\v{r}ejov, Czech Republic
\item South African Astronomical Observatory, PO Box 9, 7935 Observatory, Cape Town, South Africa
\item Southern African Large Telescope Foundation, PO Box 9, 7935 Observatory, Cape Town, South Africa
\item Special Astrophysical Observatory of the Russian Academy of Sciences, Nizhnii Arkhyz, 369167, Russia
\end{affiliations}}

\begin{abstract}

\vskip 5mm

Gravitational wave emission can lead to the coalescence of close pairs of compact objects orbiting each other\cite{hul75,abb16}. 
For the case of neutron stars such mergers may yield masses above the Tolman-Oppenheimer-Volkoff limit, leading to the formation 
of black holes\cite{abb17}. For the case of white dwarfs the merger product may exceed the Chandrasekhar limit, leading either to 
a thermonuclear explosion as Type Ia supernova\cite{ibe84,pak11}, or to a collapse forming a neutron star\cite{sai04,she12}. If a 
Type\,Ia supernova explosion is avoided, the merger of two massive white dwarfs is expected to form a hydrogen- and helium-free 
circumstellar nebula with a hot and luminous, rapidly rotating and highly magnetized central star for several 10,000\,yr before 
its final collapse\cite{ji13,sch16}. Here we report the discovery of a hot star with an emission line dominated spectrum in the 
centre of a circular mid-infrared nebula. Both the star and the nebula appear to be free of hydrogen and helium. Our tailored 
stellar atmosphere and wind models indicate a stellar surface temperature of about 200,000\,K, and a record outflow velocity of 
$16,000 \, \kms$. This extreme velocity, together with the derived mass outflow rate, imply rapid stellar rotation and a strong 
magnetic field aiding the wind acceleration. The {\it Gaia} distance\cite{bai18} of the star leads to a luminosity of $10^{4.5} 
\, \lsun$, which matches models of the post-merger evolution of super-Chandrasekhar mass white dwarfs\cite{sch16}. The high 
stellar temperature and the nebular size argue for a short remaining lifetime of the star, which will produce a bright optical 
and high-energy transient upon collapse\cite{des06}. Our observations indicate that super-Chandrasekhar mass white dwarf mergers 
can indeed avoid a thermonuclear explosion as Type\,Ia supernova, and provide empirical evidence for magnetic field generation 
in stellar mergers. 
\end{abstract}

\begin{figure}
\caption{{\bf WS35: a new mid-infrared nebula in Cassiopeia.} Upper panels: {\it WISE} 22\,$\mu$m image of the nebula at two 
intensity scales, highlighting details of its structure. The position of the central star is indicated by a circle. Bottom 
panels, from left to right: {\it WISE} 12\,$\mu$m and IPHAS H$\alpha$ images of the nebula and its central star. At the 
distance of WS35 of $\approx3$ kpc, 1 arcmin corresponds to $\approx$0.86 pc. The coordinates are in units of RA (J2000) and Dec. 
(J2000) on the horizontal and vertical scales, respectively.
\label{fig:ws35} }
\begin{center}
\includegraphics[width=12cm]{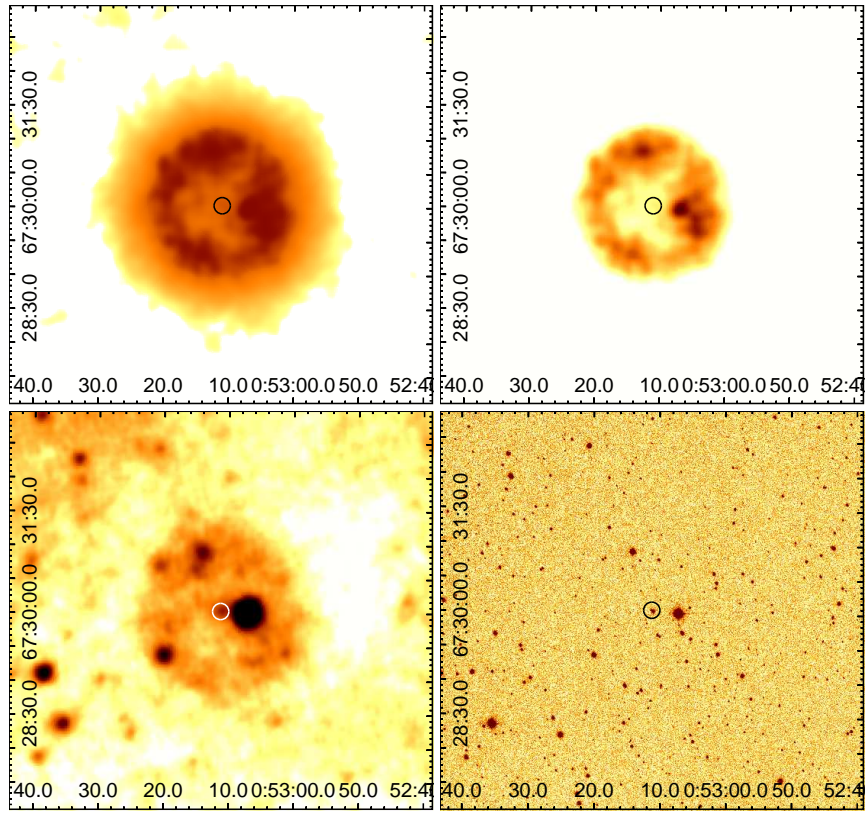}
\end{center}
\end{figure}

During our search for mid-infrared circumstellar nebulae\cite{gva10, gva12}, we discovered a new object in the constellation 
Cassiopeia (Fig.\,\ref{fig:ws35}) using data of the {\it Wide-field Infrared Survey Explorer} ({\it WISE})\cite{wri10}. 
At 22\,$\mu$m the nebula (named WS35; cf. ref.\cite{gva12}) appears as a circular shell with ragged edges and angular radius of 
$\approx75$ arcsec. The higher contrast 22\,$\mu$m image of WS35 shows a diffuse halo with a radius of $\approx110$ arcsec 
surrounding the shell. For the distance to WS35 of $\approx3$\,kpc (see below), the linear radii of the shell and halo are, 
respectively, $\approx1.1$ and 1.6 pc. The shell is also visible in the {\it WISE} 12\,$\mu$m image, where it appears as a diffuse 
circular nebula of the same angular size as the 24\,$\mu$m shell. Surprisingly, despite of the moderate extinction towards WS35 
(see below) the infrared nebula has no optical counterpart in the INT Photometric H$\alpha$ Survey of the Northern Galactic Plane 
(IPHAS)\cite{gon08} (see Fig.\,\ref{fig:ws35}). We identified the central star of WS35 with an optical star ($V\approx$15.5 mag; 
see Methods) located at right ascension 00\,h 53\,m 11.21\,s and declination +67$^{\circ}$ 30$^\prime$ 02.1$^{\prime\prime}$. 
Below, we use the  same name for the nebula and its central star.

Optical follow-up spectroscopy of WS35 with the Russian 6-m telescope (see Methods) revealed an emission-line dominated spectrum of 
the central star, reminiscent of oxygen-rich Wolf-Rayet (WO-type) stars (Fig.\,\ref{fig:spec}). However, WS35's emission lines are 
stronger and broader than those of even the most extreme WO stars. Most notably, the O\,{\sc vi} $\lambda\lambda$3811, 3834 emission 
doublet shows an equivalent width of EW(O\,{\sc vi})$\approx2300$\,\AA \, and a full width half measure of $\approx300$\,\AA. 
We note that no nebular lines are visible in the obtained long-slit spectrum.

We analysed the optical spectrum of WS35 using the Potsdam Wolf-Rayet (PoWR) code for expanding stellar atmospheres (cf. 
ref.\cite{ham04}), tailored for WS35 (Methods). In Fig.\,\ref{fig:spec}, we compare our best-fitting model with 
the observed spectrum. The line spectrum is reproduced well, except of two missing emission lines at 4340 and 6068 \AA \, which 
are likely due to line transitions of highly ionized neon (Ne\,{\sc viii}; cf. ref.\,\cite{wer07}), which are not included in the 
present modelling. Our model fit yields a stellar temperature at the base of the wind of $200,000^{+12,000}_{-22,000}$\,K, and 
a chemical composition dominated by oxygen and carbon with mass fractions of $0.8\pm0.1$ and $0.2\pm0.1$, respectively. 
Remarkably, the absence of the He\,{\sc ii} $\lambda$4686 emission line implies an essentially helium-free composition, with a 
conservative upper limit on the helium surface mass fraction of $<0.1$. From the width of the O\,{\sc vi} $\lambda\lambda$3811, 
3834 emission we derive a terminal wind velocity of $v_\infty = 16,000\pm1,000 \, \kms$. The strength of the emission lines 
is best reproduced with a mass-loss rate of $\dot{M}=(3.0\pm0.4)\times10^{-6} \myr$ (where $\msun$ is the solar mass) 
with a wind clumping factor of 10 (Methods).

Based on the distance of $3.07^{+0.34} _{-0.28}$\,kpc derived from the {\it Gaia} data release\,2 (ref.\,\cite{bai18}), our model
reproduces the observed flux distribution from ultraviolet to infrared with a luminosity of $\log(L/\lsun)=4.51\pm0.14$ (where
$\lsun$ is the solar luminosity) and with an interstellar extinction of $E(B-V)=0.8$\,mag (see Fig.\,\ref{fig:spec} and Methods). 
This luminosity is five times lower than that of even the faintest massive WO-type Wolf-Rayet stars\cite{tra15} ($\log(L/\lsun)>5.2$), 
and at least three times higher than those of the known low-mass [WO]-type central stars of planetary nebulae\cite{ges06} 
($\log(L/\lsun)<4.0$). 

Very few hot stars free of hydrogen and helium are known in the Milky Way. Besides the rare DQ white dwarfs\cite{duf07}, these are 
two hot white dwarfs with unusually high masses ($0.7\dots 1.0 \, \msun$) which have been suggested to have formed through the merger 
of two carbon-oxygen white dwarfs\cite{wer15}. Recent models for the evolution of super-Chandrasekhar mass carbon-oxygen white dwarf 
merger remnants by Schwab et al.\cite{sch16} match the properties of WS35 remarkably well. Not only does their fiducial model fit 
the Hertzsprung-Russell diagram position of WS35 (Fig.\,3). They also predict an extended episode of extreme mass loss during and 
shortly after the merger event in an expanded cool supergiant stage which will form a slowly-expanding hydrogen- and helium-free 
circumstellar nebula with an expansion velocity of about $100 \, \kms$. After the settling of the post-merger product in a hot and 
compact stage, its ultraviolet emission ionizes the surrounding nebula. Its current $\sim100$ times faster wind sweeps the preceding 
slow material into a shell, which appears inside the density-bounded \hii region, which forms the diffuse halo (see 
Fig.\,\ref{fig:ws35}). The expected hydrogen- and helium-free composition of the nebula and the high temperature of its central star 
(leading to triple ionization of oxygen) suggest that no optical nebular lines form and that the observed mid-infrared emission from 
the nebula is dominated by the [O\,{\sc iv}] 25.89\,$\mu$m and [Ne\,{\sc v}] 14.32 and 24.32\,$\mu$m lines (cf. ref.\cite{fla11}). 
This naturally explains why the nebula WS35 does neither appear in the IPHAS image nor in our long-slit spectrum.

The merging white dwarf scenario also addresses the major puzzle about WS35, namely the extreme width of its emission lines. A 
velocity of $16,000 \, \kms$ exceeds the stellar escape speed about 10 times, and is typical for supernovae but so far unheard of 
for radiation driven winds. Pure radiation driving is in fact excluded, since the wind's kinetic energy flux exceeds the total 
radiative luminosity of the star by a factor of two (Methods). The extreme velocity can, however, be explained in the framework of 
rotating magnetic wind models. Poe et al.\cite{poe89} find that a rigidly co-rotating magnetic field can enhance the speed and mass 
outflow rate of radiation driven winds by more than a factor of three, at the cost of the star's rotational energy. We find that 
a co-rotation speed of $16,000 \, \kms$ at the Alfv\'en point in WS35, where the inertia force starts to dominate over the magnetic 
forces\cite{pet13}, requires an Alfv\'en radius of about 10 stellar radii ($\sim1.5 \, \rsun$), which is achieved with a magnetic 
field strength of $\sim 10^8$\,G. Since the whole post-merger evolution is expected to last $\sim20,000$\,yr (ref.\cite{sch16}), 
it is plausible that the corresponding magnetic torques have not yet spun down WS35. 

The generation of a strong magnetic field is indeed expected in stellar mergers\cite{bel14}. 3D magneto-hydrodynamical models of 
white dwarf mergers find a magnetization of the merger product of $2\times10^8$\,G (ref.\cite{ji13}). This compares to the peak of 
the $B$-field distribution of magnetic white dwarfs of several $10^7$\,G (ref.\cite{wic00}). The observations that the mean mass of 
magnetic white dwarfs is significantly higher than that of non-magnetic ones, and that nearly none of the known magnetic white 
dwarfs have a companion star provide strong evidence for magnetic field generation by white dwarf merging\cite{wic00}. 

Adopting the escape velocity from the expanded merger remnant of $\sim100 \, \kms$ (cf. ref.\cite{she12}) and given the angular 
radius of the nebula of 1.6\,pc, we derive an expansion age of $\sim16,000$\,yr for WS35. Together with the high temperature of the 
central star, this indicates that WS35 is close to the endpoint of its post-merger evolution. Since WS35 is more luminous than the 
1.49\,M$_{\odot}$ model of Schwab et al.\cite{sch16}, it appears likely that its mass also exceeds the Chandrasekhar limit, with 
the exciting perspective that it will produce a low-mass neutron star in the near future, accompanied by a high-energy transient 
and a fast evolving supernova\cite{des06}. It is also likely that the inner shell of the circumstellar nebula will persist until 
the explosion and, given the expected low mass of supernova ejecta, will affect the appearance of the supernova remnant 
$\sim100$\,yr after explosion. 

The merging of two stars in a binary system is not a rare event. About 10 per cent of the massive main sequence stars are thought 
to be merger products\cite{min14}, as well as a similar fraction of the known white dwarfs\cite{mao18}. The very unusual wind of 
WS35 strongly supports the idea that stellar mergers can indeed produce highly magnetized stars, which would explain the magnetic 
stars of the upper main sequence\cite{fos16} and the formation of the magnetic white dwarfs\cite{wic00}. Our results also enlighten 
the ongoing debate on whether a super-Chandrasekhar mass merger of two carbon-oxygen white dwarfs leads to a Type\,Ia supernova. 
Here, WS35 appears to provide a counter example, which will likely produce a neutrino-flash and a gamma-ray burst\cite{dar92},
followed by a very fast and subluminous Type\,Ic supernova\cite{des06}. 

\section*{Bibliography}
\vspace{1cm}

\bibliographystyle{naturemag}

\begin{addendum}
\item V.V.G. acknowledges support from the Russian Science Foundation grant No. 14-12-01096. O.V.M. and A.Y.K.
  acknowledge support from the Russian Foundation for Basic Research grant 16-02-00148. G.G. acknowledges financial support 
  from the Deutsche Forschunsgemeinschaft (DFG) under grant No.\ GR\,1717/5-1. A.Y.K. also acknowledges support 
  from the National Research Foundation (NRF) of South Africa.
  \item[Author Contributions]
  V.V.G., G.G. and N.L. jointly analysed and interpreted observational data and wrote the manuscript. O.V.M. obtained and 
  reduced the spectroscopic material. A.Y.K. provided ideas for interpretation of the nebula. A.S.M. and O.I.S. obtained 
  optical photometry. All authors discussed the results and commented on the manuscript.
  \item[Author Information] The authors declare no competing financial interests. Correspondence and requests for materials 
  should be addressed to V.V.G. (vgvaram@mx.iki.rssi.ru).
\end{addendum}

\newpage

\newpage

\begin{figure}
\caption{{\bf Spectral modelling of WS35.} Upper panel: Observed flux distribution of WS35 in absolute units, including the 
calibrated spectrum (black line) and photometric measurements (blue rectangles), compared to the emergent flux of the model 
continuum (red line), in the optical also shown with lines. The model flux has been reddened and scaled to the distance 
according to the parameters given in Table\,1. Bottom panel: Observed optical spectrum of WS35 (black line), compared with 
our best-fitting model (red line) with the parameters as given in Table\,1. The lines fitted by the model are highlighted. 
\label{fig:spec} }
\begin{center}
\includegraphics[width=22cm,clip]{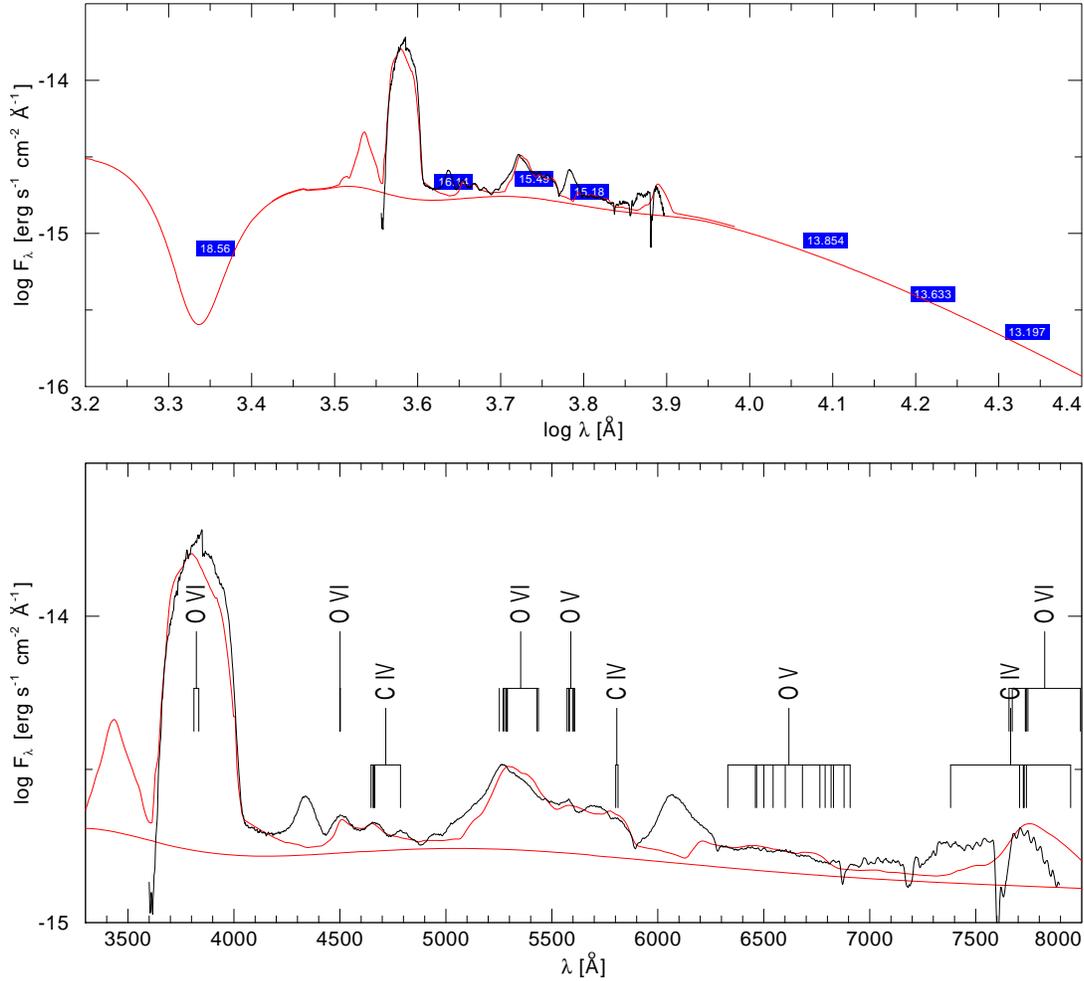}
\end{center}
\end{figure}

\newpage

\begin{figure}
\caption{{\bf Hertzsprung-Russell diagram position of WS35.} The cross marks the position, with error bars, of WS35. It is 
compared to the evolutionary track of the fiducial carbon-oxygen white dwarf post-merger model of Schwab et al.\cite{sch16}, 
which starts at the black dot, and where the model has spent about 100, 5,000, 6,000 and 16,000\,yr after the merger at the 
coloured dots labelled 1, 2, 3 and 4, respectively.
\label{fig:sch} }
\begin{center}
\includegraphics[width=10cm,clip]{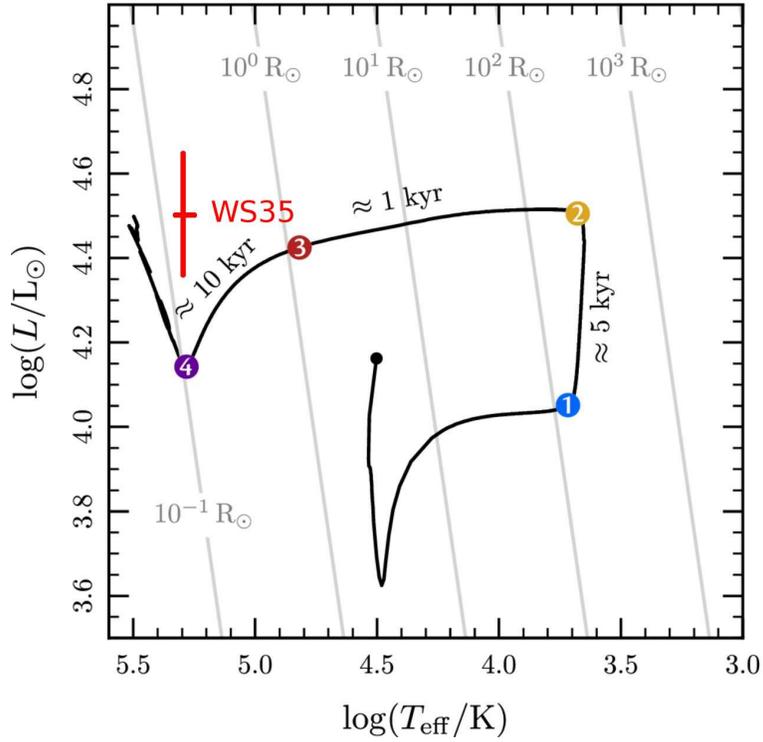}
\end{center}
\end{figure}

\newpage

\begin{table}
\caption{Stellar parameters and surface abundances for WS35.}
\label{tab:par} 
  \begin{center}
  \begin{tabular}{ll} 
    \hline  
    $\log(L_*/\lsun)$ & $4.51\pm0.14$ \\ 
    $T_*$ (K) & $200,000^{+12,000}_{-22,000}$  \\
    $R_*$ ($\rsun$) & $0.151^{+0.035}_{-0.013}$ \\ 
    $\dot{M}$ ($\myr$) & ($3.0\pm 0.4) \times 10^{-6}$ \\ 
    $D$ & $10$ \\
    $v_\infty$ ($\kms$) & $16,000\pm1,000$ \\ 
    $v_{\rm D}$ ($\kms$) & $250$ \\
    $\beta$ & $1.0$ \\
    $d$ (kpc) & $3.07^{+0.34} _{-0.28}$ \\
    $E(B-V)$ (mag) & 0.8 \\
    $R_V$ & 3.1 \\
    He (mass fraction) & $< 0.1$ \\
    C (mass fraction)  & $0.2\pm0.1$ \\
    O (mass fraction)  & $0.8\pm0.1$ \\
    Fe-group (mass fraction) & $1.6\times10^{-3}$ \\
    \hline 
  \end{tabular}
  \end{center}
\end{table} 

\newpage

\begin{methods}

\subsection{Observations.}
%
WS35 was observed with the 6-m telescope of the Special Astrophysical Observatory of the Russian Academy of Science 
(SAO RAS) using the Spectral Camera with Optical Reducer for Photometric and Interferometric Observations 
(SCORPIO)\cite{afa05} on 2017 July 20. The spectrum was obtained in the long-slit mode with a slit width of 1 arcsec, 
using the grism VRHG550G. This spectral setup covered a wavelength range of $\approx3600-7800$ \AA \, with a spectral 
resolution of 12 \AA \, (estimated using spectra of a He-Ar-Ne lamp). One exposure of 30 min was taken with a seeing of 
$\approx1.5$ arcsec. For flux calibration the spectrophotometric standard star BD+33$^\circ$2642 (ref.\cite{oke90}) was observed.
The spectrum was reduced using the {\sc score} package written by O.~Maryeva and P.~Abolmasov in the IDL language for 
SCORPIO long-slit data reduction\cite{mar}. The package incorporates all the standard stages of long-slit data reduction 
process.

We searched for optical photometry of WS35 using the VizieR catalogue access tool\cite{viz} and found that the existing
data are very different from each other. We therefore decided to obtain a contemporary photometry of WS35 and to search 
for possible variability of this star. WS35 was observed in the $B, V$ and $R_{\rm c}$ filters using the CCD photometer 
of the 1-m Zeiss-1000 telescope of the SAO RAS. A total of nine measurements were carried out in the period from 2017 August 
18 to 2018 April 30. We found that in all three filters the brightness of WS35 was fairly stable with mean magnitudes of 
$B=16.14\pm0.02$, $V=15.49\pm0.01$ and $R_{\rm c}=15.18\pm0.01$. We used these values for modelling the spectral energy 
distribution (SED) of WS35 (see next Section).

\subsection{Spectral analysis.} \label{sec:spec}
%
WS35 displays an extremely early WO-type emission-line spectrum, implying a very high temperature, and a carbon- and oxygen 
dominated surface composition. Based on the equivalent widths of the primary classification lines O\,{\sc vi} 
$\lambda\lambda$3811, 3834 (with EW(O\,{\sc vi})$\approx2300$\,\AA) and O\,{\sc v} $\lambda$5590 (with EW(O\,{\sc v})$\leq120$ 
\AA), we derive a ratio of $\log$(EW(O\,{\sc vi})/EW(O\,{\sc v}))$\geq1.3$ putting WS35 in the earliest WO sub-class WO1 
(ref.\,\cite{cro98}).

We performed a quantitative spectroscopic analysis of WS35 using non-LTE model atmospheres for Wolf-Rayet stars as described in
refs\cite{koe02,gra02,ham03}. The models include the relevant ions of He, C, O, and the Fe-group. In particular, these are the 
first models for WO stars that include the important opacities of the Fe M-shell ions (Fe\,{\sc ix}--{\sc xvi}; cf. 
ref.\cite{gra05}) which are responsible for the Fe-opacity peak arising near temperatures of $\sim150,000$\,K\cite{igl96}. We 
extended the present models to include the continuum opacities of the high ionization stages C\,{\sc v} and O\,{\sc vii}, to 
prevent the models from becoming translucent in the X-ray range. We assumed a clumped wind structure and a $\beta$-type velocity 
law with a fixed acceleration parameter $\beta=1$. Furthermore, we used line broadening with a fixed Doppler broadening velocity 
of $v_{\rm D}=250 \, \kms$, and used a fixed abundance of Fe-group elements corresponding to a Galactic (relative) metal 
distribution\cite{gra02}.

The SED was fit using the extinction law of Cardelli et al.\cite{car88} with the total-to-selective absorption ratio of
$R_V=3.1$. The observed SED was constructed from the flux calibrated spectrum and photometric data. Besides our own optical 
photometry, we used the $U$ magnitude of $18.56\pm0.05$ from the Mikulski Archive for Space Telescopes (MAST)\cite{mik} and 
the $(J, H, K_s)$ magnitudes of ($13.85\pm0.03,13.63\pm0.04,13.20\pm0.04$) from the Two-Micron All Sky Survey (2MASS) All-Sky 
Catalog of Point Sources\cite{cut03,skr06}. Because the ultraviolet to optical flux distribution is substantially affected by 
interstellar extinction, the derived luminosity of WS35 relies mainly on the continuum-fit in near-infrared range (cf. 
Fig.\,\ref{fig:spec}). To take the influence of infrared emission lines into account, we place the continuum slightly below 
the photometric fluxes. The uncertainty of $\sim0.1$\,dex in this fitting procedure is the main error source for the derived 
luminosity of WS35.

The wind velocity of $v_\infty =16,000\pm1000 \, \kms$ is predominantly determined from the width of the strong 
O\,{\sc vi}\,$\lambda$3811--34 line. This value is further backed up by the width of the O\,{\sc v}\,$\lambda$5590 line. 
In our models oxygen shows a pronounced onion-shell structure where O\,{\sc v} is the leading ion in the outer layers, 
followed by O\,{\sc vi} and O\,{\sc vii} further inward. As a consequence O\,{\sc v}\,$\lambda$5590 shows a very broad, 
almost rectangular flat-top line profile, as it is expected for optically-thin emission from a hollow shell. While the
blue part of this line profile is blended with O\,{\sc  vi}\,$\lambda$5300, its red edge is clearly visible near 5900\,\AA,
supporting the high value of $v_\infty$ derived in this work. If helium is included in our models, He\,{\sc ii}\,$\lambda$4686 
displays a similar, broad flat-top line profile, which is absent in the observed spectrum of WS35. According to our models 
this line profile would be clearly visible for He mass fractions $\gtrsim 0.1$.

The relative abundance of C and O was constrained by the relative strength of the C\,{\sc iv} and O\,{\sc vi} features in 
the line blend around 4700\,\AA. The mass-loss rate $\dot{M}$ was constrained by the strength of the O\,{\sc v}\,$\lambda$5590 
line, which reacts very sensitively to changes in the wind density. However, we note that there is a degeneracy between 
$\dot{M}$ and the wind clumping-factor $D$ with $\dot{M}\sqrt{D}$=const, where $D$ can only be determined very roughly from the 
strength of the electron-scattering wings of strong emission lines (cf. ref.\cite{ham98}). For WS35 the red wing of the 
O\,{\sc vi}\,$\lambda$3811--34 emission implies a value of $D\gtrsim 10$. In our analysis, we adopt $D=10$ which is a typical 
value found in studies of emission-line stars (e.g. refs\cite{bes14, tra15}). But, the uncertainties are large and may affect 
the derived mass-loss rate by up to a factor two. Finally, the stellar temperature $T_*$ was constrained from the strength and 
profile shape of the O\,{\sc vi}\,$\lambda$3811--34 and O\,{\sc vi}\,$\lambda$5300 lines. Our model fit with $T_* =200,000$\,K 
reproduces both the strength and the profile shape well. For lower $T_*$ the strength of the O\,{\sc vi}\,$\lambda$3811--34 
emission becomes weaker. For higher $T_*$ the line profiles, in particular the one of the O\,{\sc vi}\,$\lambda$3811--34 line, 
change towards a flat-top profile instead of the observed triangular ones. The reason is that for hotter temperatures these lines 
are formed further outwards in the wind, at velocities close to $v_\infty$.

\subsection{Wind velocity of WS35.} \label{sec:vel}
%
The extremely high wind velocity of WS35 ($v_\infty =16,000 \, \kms$) is very unusual for Wolf-Rayet type winds. 
Gr\"afener \& Vink\cite{gra13} found typical ratios of $v_\infty /v_{\rm esc} \sim 1.3$, where $v_{\rm esc}$ is the escape 
velocity, for WC/WO stars. Such a fixed ratio is also supported by the theory of optically-thick radiatively-driven 
winds\cite{gra17}. For WS35 this would imply an escape velocity of $12,000 \, \kms$ or slightly lower if we consider that 
the few WO stars in the sample of Gr\"afener \& Vink\cite{gra13} showed slightly higher ratios of $v_\infty$/$v_{\rm esc}$.

If we adopt a mass of the order of $M_*=1.5 \, \msun$ for WS35 (see main text), this value would imply a stellar radius 
of $R_* =2GM/v_{\rm esc} ^2 \gtrsim 0.004 \, \rsun$ which is a factor of 40 lower than the value derived in our analysis 
($\approx0.15 \, \rsun$). This discrepancy indicates that a different wind driving mechanism than for ordinary Wolf-Rayet 
stars may be in action, i.e. the wind of WS35 may not be driven by radiation pressure alone.

This conclusion is further supported by the high mechanical wind luminosity of WS35 $L_{\rm wind}=\dot{M}(v_\infty^2 /2 
+M_*G/R_*)=6\times10^4 \, \lsun$. This value is two times higher than the radiative luminosity $L_*$ that we derived in our 
analysis, i.e. the wind energy exceeds the maximum limit of $L_{\rm wind}= L_*$ that can be achieved by radiative driving. 
Even considering the uncertainties in $\dot{M}$ due to wind clumping, this makes it very unlikely that the wind of WS35 is 
radiatively driven. In fact, the radiative acceleration computed within our wind models provides only 5 per cent of the work 
that is necessary to drive the wind, a value that is remarkably small compared to previous studies\cite{gra02,gra05} of the 
wind dynamics of WC stars where a consistency within 50 per cent was achieved. Therefore, a magnetically supported, rotating 
wind appears required to explain the high wind velocity observed for WS35.

\subsection{Data availability.} \label{sec:dat}
%
Data available upon request from the corresponding author.

\end{methods}

\end{document}